\documentclass[aps,a4paper,twocolumn,floatfix,showpacs,
unsortedaddress]{revtex4}
\usepackage{bm,amsmath,amssymb,amsfonts}

\usepackage[dvips]{graphicx}
\usepackage[colorlinks=true,citecolor=green,urlcolor=red,filecolor=blue]{hyperref}
\usepackage{color}
\definecolor{mycolor}{rgb}{0.4,0.10,0.8}
\begin{document}

\preprint{APS/123-QED}
	
\title{Anomalous response, re-emerging localization-transition and robust corner sates in self-similar peano fractal}% Force line breaks with \\

\author{Rhiddha Acharjee}
\email{rhiddhaacharjee777@gmail.com}
\affiliation{
    Department of Physics, University of Calcutta,
    92, Acharya Prafulla Chandra Road, Kolkata,
    West Bengal-700009, India}
\author{Sayan Bhattacharya}
\email{sayanbhatta2002@gmail.com}
\affiliation{
    Department of Physics, Acharya Prafulla Chandra College,
    New Barrackpore, Kolkata, West Bengal-700131, India}
\author{Atanu Nandy}
\email{atanunandy1989@gmail.com}
\affiliation{
    Department of Physics, Acharya Prafulla Chandra College,
    New Barrackpore, Kolkata, West Bengal-700131, India}
\author{Debnarayan Jana}
\email{djphy@caluniv.ac.in }
\affiliation{
    Department of Physics, University of Calcutta,
    92, Acharya Prafulla Chandra Road, Kolkata,
    West Bengal-700009, India}
\begin{abstract}
	Anomalous spectral response from the non-interacting electrons on a peano fractal is recorded within tight-binding framework. It is observed that a minimal off-diagonal anisotropy can modulate the transport characteristics of the lattice. Drastic modification in the nature of the eigenstates has been validated through the workout of level spacing statistics. Further, a two-dimensional quasiperiodic Aubry-Andr\'e and Harper variation in the overlap integral can invite re-entrant localization-delocalization transition for the self-similar model. This is associated with the possible existence of single-particle mobility edge in the energy landscape. Our results are corroborated by the evaluation of inverse participation ratio, fractal dimension and quantum dynamical response. The spectrum is shown to formulate a quantum butterfly pattern with the frequency of the modulation. Chiral symmetry protected edge states are found to be insensitive with respect to moderate diagonal disorder.
\end{abstract}
%\keywords{Localization-transition, fractal, quantum dynamics, edge modes.}
\maketitle
%%%%%%%%%%%%%%%%%%%%%%%%%%%%%%%%%%%%%%%%%%%%%%%%%%%%%%%%%%%%%%%%%%%%%%%%%%%%%%%%%%
\section{Introduction}
\label{intro}
%%%%%%%%%%%%%%%%%%%%%%%%%%%%%%%%%%%%%%%%%%%%%%%%%%%%%%%%%%%
%%%%%%%%%%%%%%%%%%%%%%%%%%%%%%%%%%%%%%%%%%%%%%%%%%%%%%%%%%%
\begin{figure*}[ht]
\centering
(a) \includegraphics[clip,width=5.0 cm,angle=0]{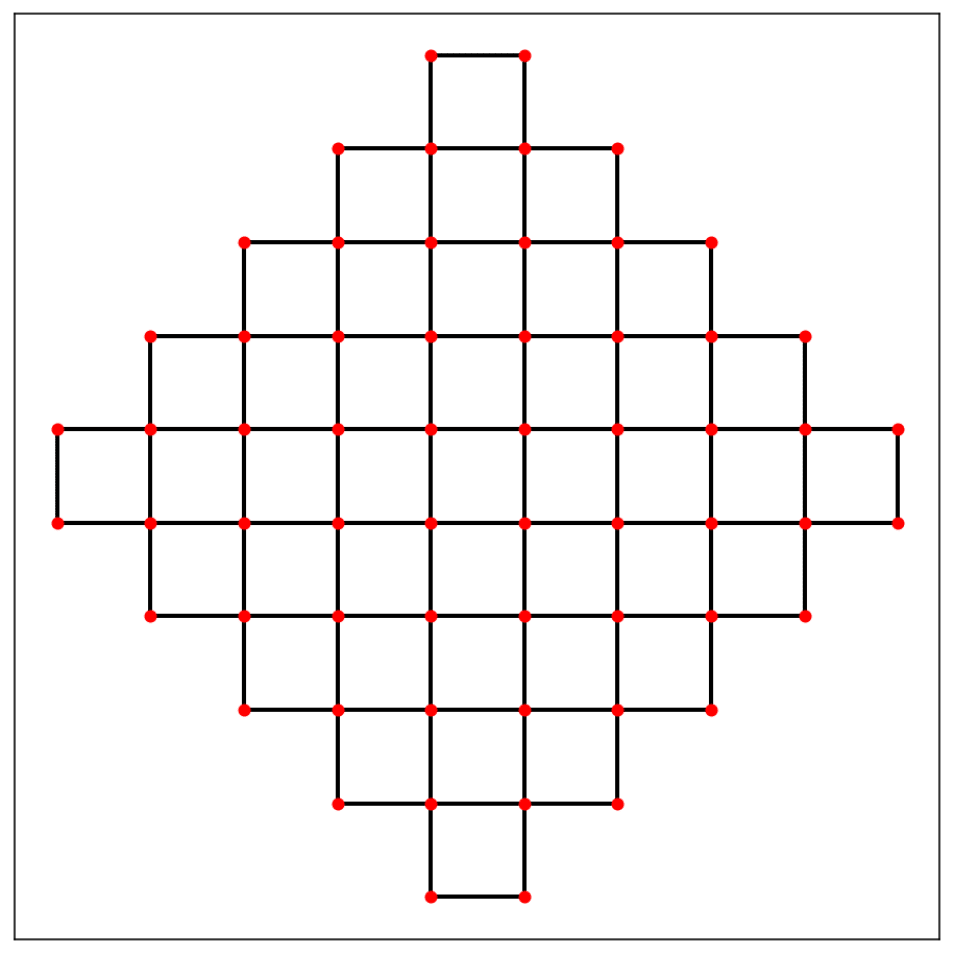}
(b) \includegraphics[clip,width=5.0 cm,angle=0]{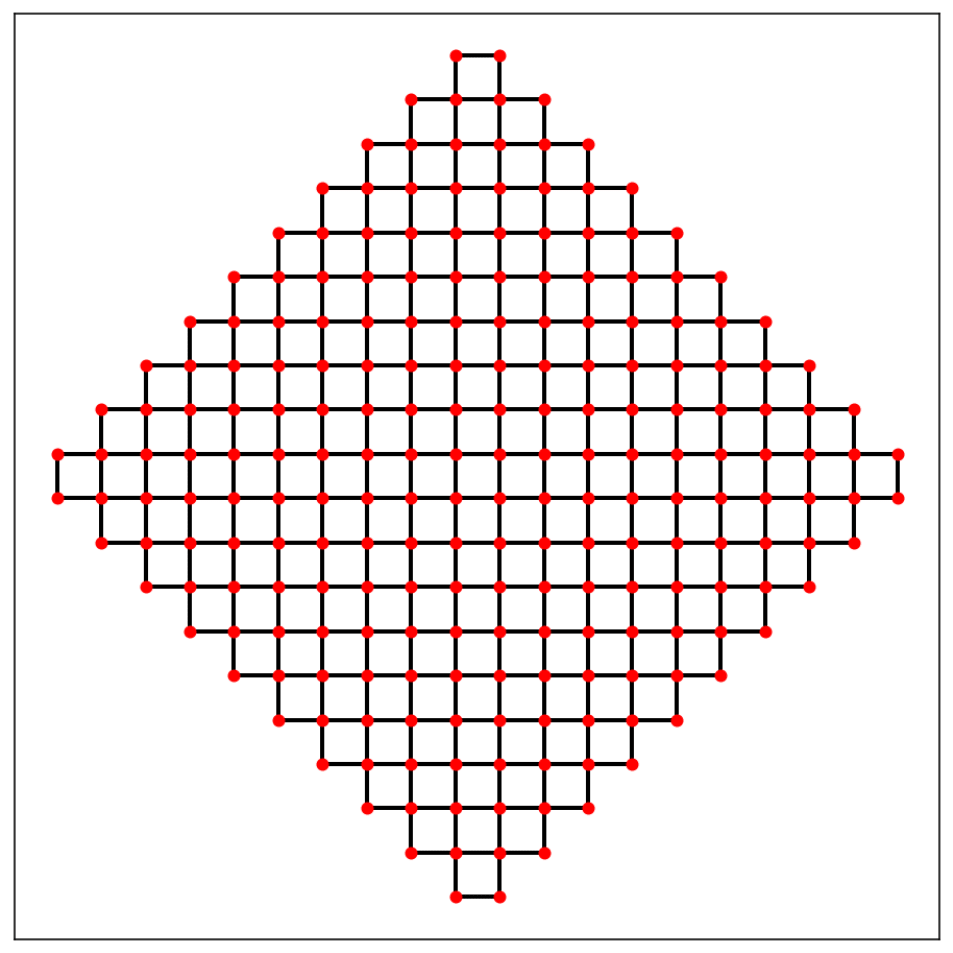}
(c) \includegraphics[clip,width=5.0 cm,angle=0]{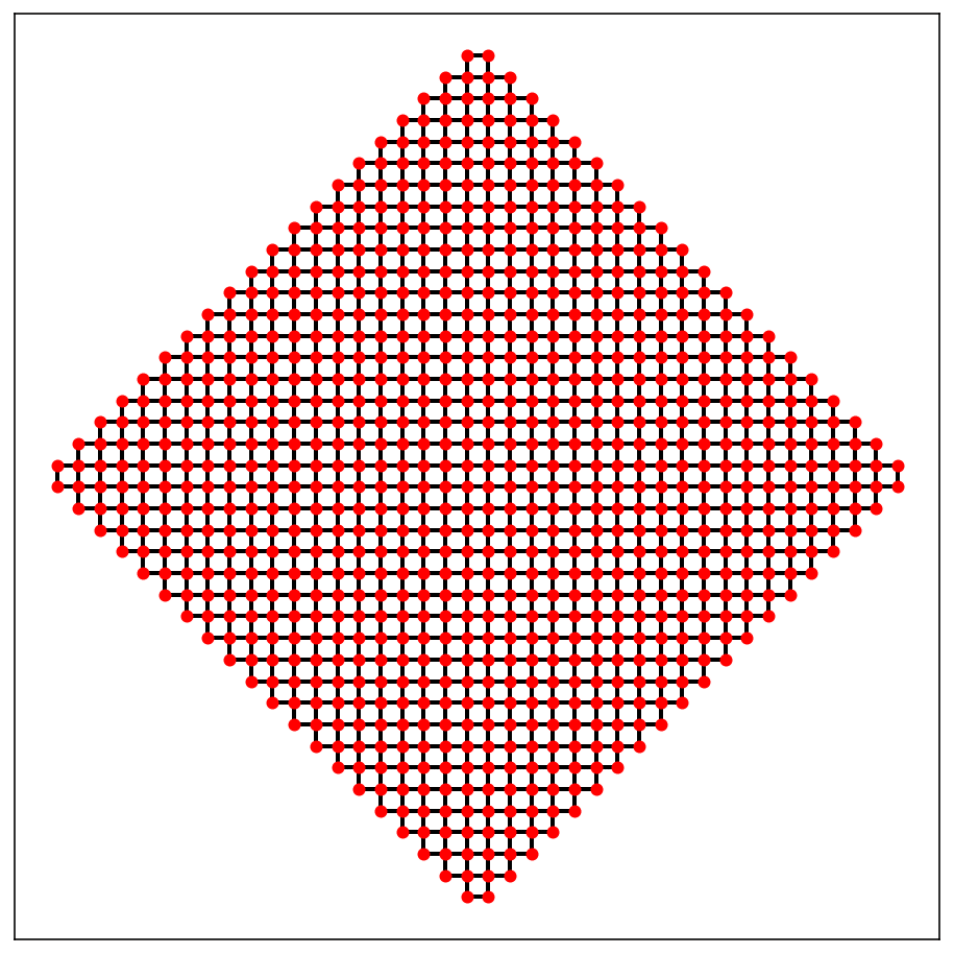}

\caption{(Color online) Hierarchical growth pattern of the self-similar peano fractal is indicated by the three successive generations with number of sites $60$, $220$ and $860$ respectively.}  
\label{sys}
\end{figure*}
%%%%%%%%%%%%%%%%%%%%%%%%%%%%%%%%%%%%%%%%%%%%%%%%%%%%%%%%%%%
Advancement in the engineering of different quantum systems have triggered quantum technology applications including the inevitable field of quantum simulation. Because of the significant impact over diverse fields, a number of experimental platforms~\cite{seba1,rod,seba2,seba3} explore different interesting strategies for the fabrication and control of completely artificial networks. These systems include geometries that are beyond general homogeneous Euclidean objects ranging from fractal entity to quasiperiodic systems. Those systems can be generated by molecular assembly~\cite{shang}, or with the aid of scanning tunneling microscopy~\cite{kemp} photonic network made of coupled optical fibres~\cite{xu,t} or cold atoms in the optical tweezers. 

Fractal geometries~\cite{RojoFrancas2024,an2,bp1,bp2,bp3,wang,ac,ac2,sayan,eek} are not translationally invariant systems but they are characterized by an inherent \textit{self-similarity} giving rise to fractional Hausdorff dimension. This typical structural non-repetitive pattern unfolds a fascinating domain of study of tunable localization of single-particle eigenstates which has received considerable momentum in recent times. The non-trivial spectral response of this system makes the experimental realization of such systems extremely challenging and interesting as well~\citep{RojoFrancas2024,xie,yang}. Moreover, the topologically protected edge states governed by the consequence of bulk-edge correspondence are found to be present where there is well-determined bulk of the system. But the sustainability of edge modes in fractal structures, where a prevailing sense of bulk of the system is absent, is a challenging aspect which has been recently demonstrated experimentally using array of photonic waveguides. It is needless to mention that the standard femto-second laser writing formalism has made substantial contribution in this regard. Recent further studies also reveal that topological modes can appear in hierarchical fractal lattices~\cite{Lage2025,Li2022,Chen2023,He2024,Li2023,Lai2024}. All these findings challenge the common understanding of bulk-edge correspondence.

Uncorrelated (random) disorder induced localization of wavefunction, originally proposed 
by P. W. Anderson~\cite{anderson}, predicts the metal-insulator transition as a consequence of the wave interference resulted from the scattered wavefunction in presence of impurities. This phase transformation is constrained to higher dimensional systems. Interestingly, the similar physics can be achieved even in one dimension if we go for a quasiperiodic potential profile of scatterers. The preliminary example is the Aubry-Andr\'{e} (AA) model, a cornerstone in the study of quasicrystals~\cite{aubry}, creates a monotonic localization transition in the spectrum with respect to the strength of disorder. The more tunable situation can be obtained with staggered or disordered environment under specific correlation of the parameters. This means at least a subset of localized modes become delocalized again with certain combination of parameters. This re-emerging transformation is generally limited to observe in the one-dimensional systems~\cite{basu,skm,zigzag} having translational periodicity. Extension of one-dimensional AAH model to its two-dimensional (2D) analogue~\cite{yu} reveals the appearance of higher-order topological phases with robust corner modes. However, controlling localization-transition working on a fractal geometry using the 2D AAH modulation remains almost unsolved issue in this regard. 

In the present article, we try to examine the tunable localization-delocalization transition for a self-similar fractal system. The spectrum is dependent on the hierarchy of the lattice as well as the quasiperiodic kinetic modulation included. Hence the transmission for such system is extremely sensitive to the abovementioned factors. To the best of our knowledge, this kind of observation has not been verified for any fractal systems. Here, we undertake a tight-binding peano fractal~\cite{peano} and address the atypical spectral feature and anomalous transport property of it induced by off-diagonal modulation. Recursively grown peano fractal is deterministic, but lacks translational ordering. Production of delocalized eigenmodes in this lattice is quite non-trivial, as expected.  However, minimal off-diagonal modulation can destroy this delocalization. As a result, the appearance of localization-delocalization transition in the spectrum for such lattice is a challenging topic that maybe observed. At this point, we look into the possibility and mechanism of the re-entrant phase transformation with nominal modulation of the kinetic parameters. As it is already reported that a subtle correlation between the parameters for a hierarchical fractal structure may host delocalized eigenfunctions~\cite{sougata,sougata2}. Specific resonant value of the parameters can turn the fragmented spectrum into an absolutely continuous one. The external magnetic perturbation may also influence the caging of wavepacket in such hierarchical networks~\cite{sayan} in an interesting fashion. Unlike to these extreme cases, we are trying to find a scenario where one can have a comprehensive control over the parameters by which we can visualize a re-emerging localization-delocalization crossover. Based on that perspective we first investigate the modification in spectral landscape of the peano fractal by incorporating minimal degree of anisotropy in the hopping integral. The remarkable change in the response prompts us to include quasiperiodic variation in the kinetic parameter which will ultimately lead to re-entrant \textit{localization-delocalization} transition hosting an obvious single-particle mobility edge (SPME) in the energy spectrum. The spectrum is reported to give rise to a quantum butterfly pattern with respect to the frequency of the modulation imposed encompassing the existence of in-gap edge modes which are found to be protected by chiral symmetry associated with the underlying system. The motivation comes from the recent progress in the context of quantum gases in optical lattices where the localization transition and the possible existence of SPME has been tested for quasiperiodic systems~\cite{Boers2007,Li2017} with the experimental realizations~\cite{Luschen2018,An2018,An2021} inclusive. 

In what follows, we summarize our findings. Sec.~\ref{model} discusses the tight-binding description of the 
fractal. Sec.~\ref{spectrum} highlights the response of the system with respect to minimal hopping modulation. Localization transition is analyzed in the Sec.~\ref{transition} and multifractal eigenspectrum
is discussed in the Sec.~\ref{butterfly}. Finally, we draw our closing remarks in the Sec.~\ref{conclude}.

\section{Model System and Hamiltonian}
\label{model}
We first try to achieve the general insight of the electronic spectrum of the hierarchical peano fractal lattice. The system can be described by the tight-binding Hamiltonian, written in the Wannier basis, as, 
	
	$$
	H=\sum_{\langle i,j\rangle}
	\left(t_{ij}\, b_i^{\dagger} b_j+\mathrm{h.c.}\right)
	+\sum_i \epsilon_i\, b_i^{\dagger} b_i .
	$$
	
	Here, $b_i$ ($b_i^{\dagger}$) denotes the local annihilation (creation) operator. $\epsilon_i$ is the on-site energy of an electron at the $i$-th site, and $t_{ij}$ is the nearest-neighbor hopping integral between the $i$-th and $j$-th sites.We introduce an off-diagonal anisotropy index $\alpha=\frac{t_y}{t_x},$ where $t_x$ and $t_y$ are the hopping integrals along the bonds connecting two nearest-neighboring sites parallel to the $x$- and $y$-axis, respectively. We shall first present our results to understand the impact of this minimal kinetic anisotropy with respect to the variation of $\alpha$. Without any loss of generality, the on-site energy of all the atomic sites is taken to be $\epsilon = 0$ throughout the numerical analysis. This numerical choice merely sets the reference of the energy spectrum. Fractal system, in general, exhibits fragmented cantor-set like eigenspectrum. Here the isotropic fractal offers delocalized egenfunctions and the anisotropy can partially destroy the extended zone, by including a non-transmitting section of allowed energy. This is encoded by the evaluation of quantum transport and level spacing distribution.
		
		With this tight-binding description we further show that quasiperiodic kinetic modulation imposed in the anisotropic peano fractal may influence the imprisonment of the excitation in an interesting controlled manner. It has already been reported that while the scale-invariant structure offers fragmented electronic spectrum, in general, a subtle combination can produce diffusive eigenstates of electron. In a different context, the simultaneity of two different quantum phases can also be achieved by incorporating quasiperiodic disorder in the parameters of a periodic system~\cite{basu}. In our discussion, the coexistence of extended and localized phases in the spectrum with respect to the parametric variation leads to a re-entrant localization-delocalization transition for this non-translationally invariant fractal geometry. This transition is primarily associated with a single-particle mobility edge in our self-similar structure.
		
		To scrutinize the nature of single-particle eigenstates we rely on the evaluation of inverse participation ratio (IPR) and normalized participation ratio (NPR) which are respectively defined for n-th eigenstate $|{\psi_n}\rangle$ as,
		
		\begin{equation}
			\mathrm{IPR}_n
			=
			\sum_{k=1}^{L}
			\left|\psi_k^n\right|^4,
			\qquad
			\mathrm{NPR}_n
			=
			\left(
			L\sum_{k=1}^{L}
			\left|\psi_k^n\right|^4
			\right)^{-1}.
		\end{equation}
		
		As we go close to thermodynamic limit, the nature of the state can be understood using the limiting values of the above quantities as demonstrated in the following table.
		
\begin{table}[h]
	\centering
	\begin{tabular}{ccc}
%		\toprule
		 $\textbf{Phase}$ & $\mathrm{IPR}$ & $\mathrm{NPR}$ \\
%		\midrule
		Extended & $O\left(\frac{1}{L}\right)$ & $1$ \\
		Localized & $O(1)$ & $0$ \\
%		\bottomrule
	\end{tabular}
\end{table}
		Average IPR and NPR are the additional quantitative measure to validate the findings which are given as,
		
		\begin{equation}
			\langle \mathrm{IPR}\rangle
			=
			\frac{1}{L}
			\sum_{n=1}^{L}
			\mathrm{IPR}_n,
			\qquad
			\langle \mathrm{NPR}\rangle
			=
			\frac{1}{L}
			\sum_{n=1}^{L}
			\mathrm{NPR}_n.
		\end{equation}
		
		$\langle \mathrm{IPR}\rangle \rightarrow 0$ and
		$\langle \mathrm{NPR}\rangle \rightarrow 1$ for an extended eigenstate, in the asymptotic limit and vice-versa for localized mode. For the critical phase, both the quantities may assume intermediate finite values.
		
		The dependence of $\mathrm{IPR}_n$ over the system size for different types of eigenstates can readily be manifested via the evaluation of fractal dimension $D_n$ as,
		
		\begin{equation}
			D_n
			=
			-\lim_{L\rightarrow\infty}
			\frac{\ln(\mathrm{IPR}_n)}
			{\ln L}.
		\end{equation}
		
		$D_n$, which corresponds to the slope of the $\ln \mathrm{IPR}_n \leftrightarrow \ln N$ plot, continuously approaches towards unity for extended mode, zero for localized mode and $0<D_n<1$ for critical state.
		
\section{Spectral properties}
\label{spectrum}
\subsection{Electronic eigenspectrum}
\begin{figure}
	\centering
%	(a)\includegraphics[width=0.35\textwidth]{levelspacingextended.png}
%	(b)\includegraphics[width=0.35\textwidth]{levelspacinglocalized.png}
	\includegraphics[width=0.35\textwidth]{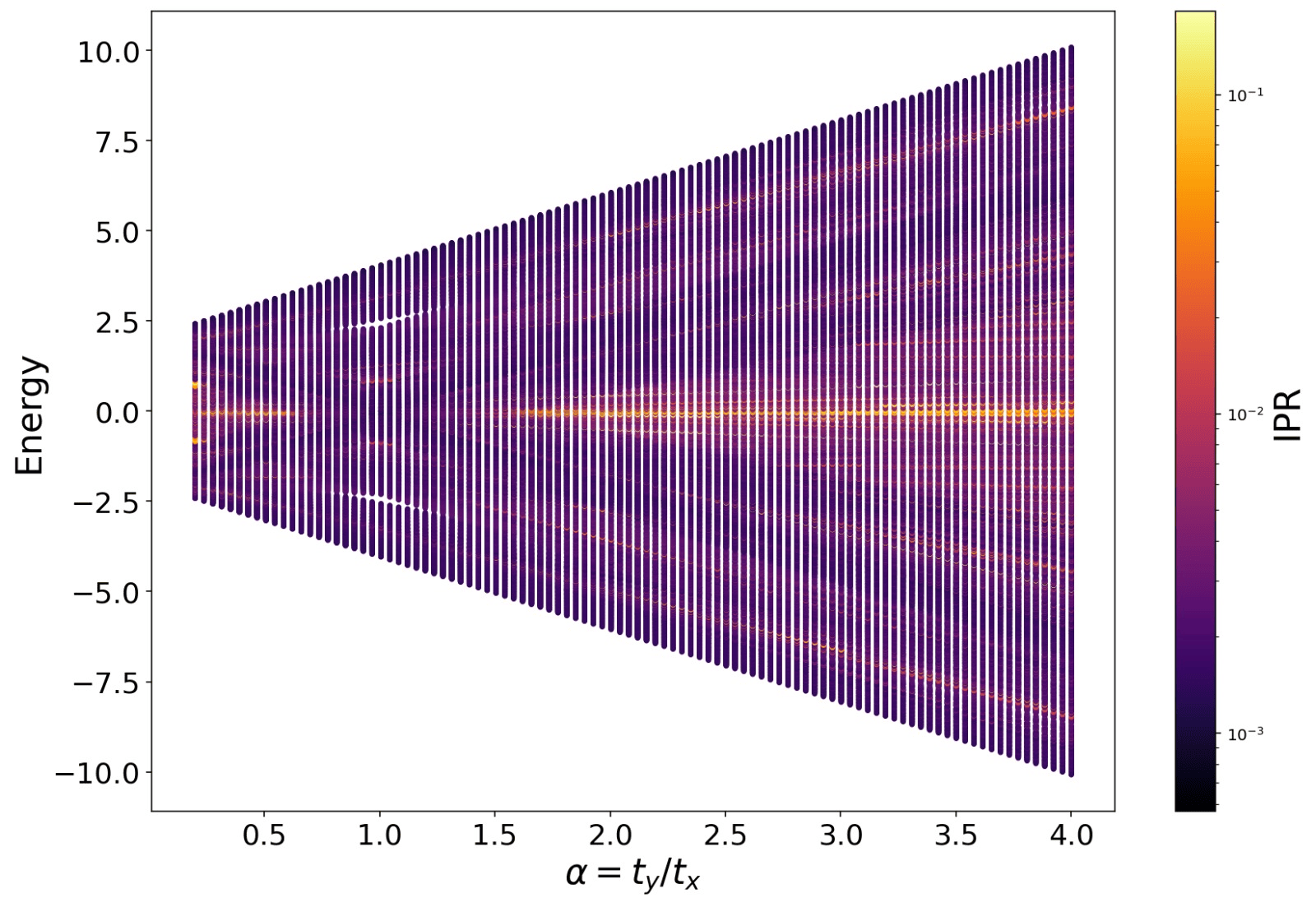}
	\caption{Distribution of allowed eigenvalues with respect to off-diagonal anisotropy index $\alpha$ in respect of their IPR values.}
	\label{spec}
\end{figure}
The evaluation of eigenspectrum is a very straightforward formalism through the calculation of the fractal dimension $D_n$ with respect to the variation of anisotropy index $\alpha$.
This fractal dimension essentially declares the nature of single-particle eigenstates.
 Here the evaluation has been made for a $5^{th}$ generation peano fractal containing $3280$ number of sites. The result is plotted in the Fig.~\ref{spec}. We see that the presence of off-diagonal anisotropy induces notable modification in the nature of single-particle eigenstates. The conducting regime bifurcates into two subsects clearly producing a zone of critical states for $\alpha \neq 1$. This plot immediately suggests the existence of critical phases in the spectrum with this minimal modulation. The anisotropy thus makes the system less accessible to the injected projectile by controlling the wave interference resulted from the network. Moreover, we can eventually conjecture that the off-diagonal anisotropy may lead to the possibility of metal-to-insulator transition for $\alpha \neq 1$.

\subsection{Transmission co-efficient}
%%%%%%%%%%%%%%%%%%%%%%%%%%%%%%%%%%%%%%%%%%%%%%%%%%%%%%%%%%%
%%%%%%%%%%%%%%%%%%%%%%%%%%%%%%%%%%%%%%%%%%%%%%%%%%%%%%%%%%%
\begin{figure}[ht]
\centering
(a) \includegraphics[clip,width=5 cm,angle=0]{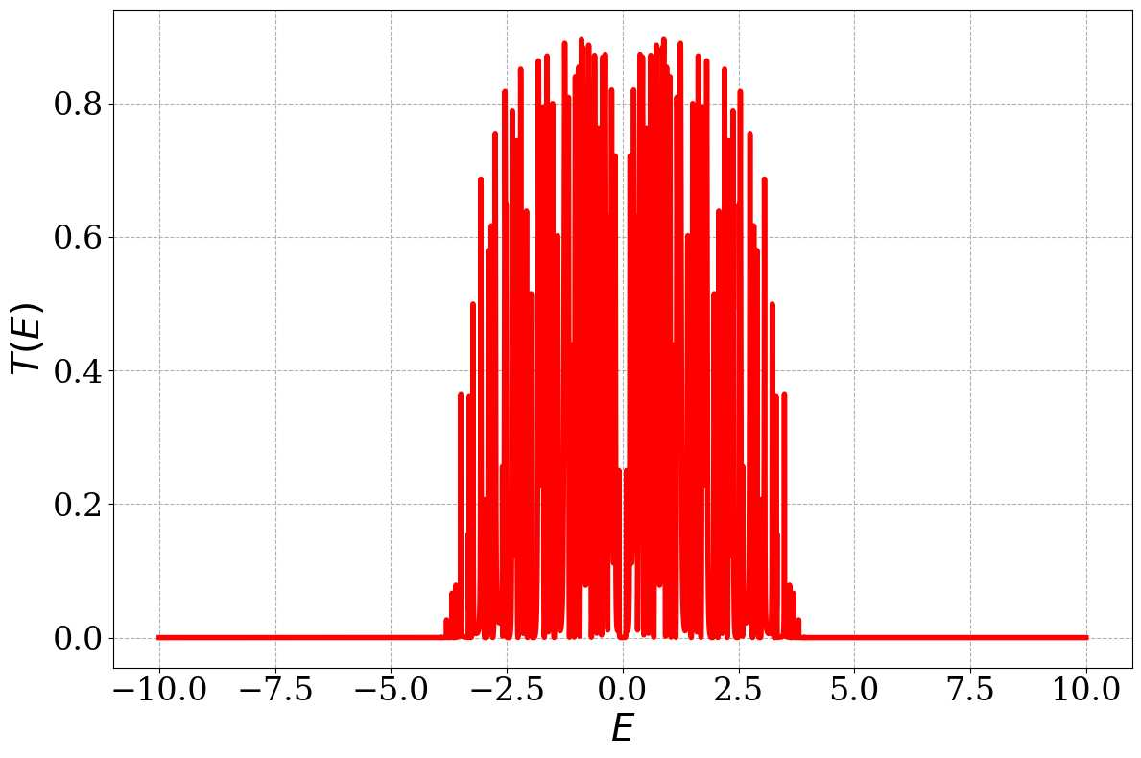}\\
(b) \includegraphics[clip,width=5 cm,angle=0]{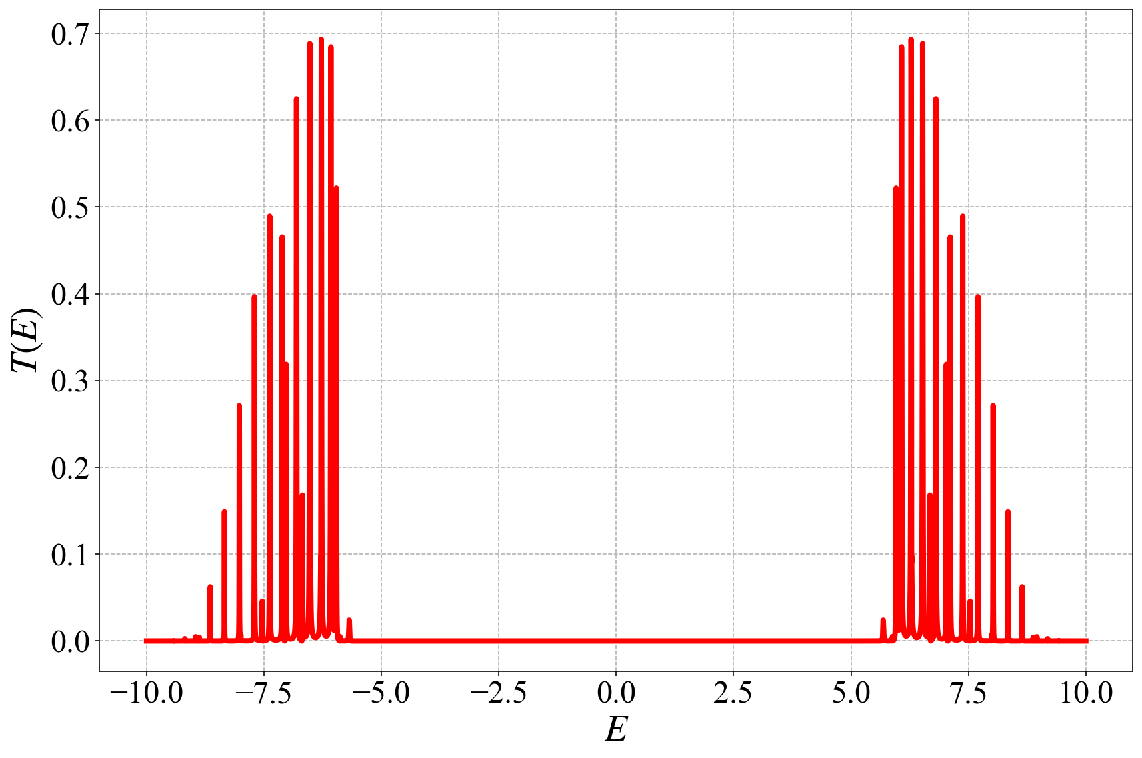}
%(c) \includegraphics[clip,width=5.3 cm,angle=0]{iprpeano.png}
%(d) \includegraphics[clip,width=5.3 cm,angle=0]{iprpeanoans.png}
%(e) \includegraphics[clip,width=5.3 cm,angle=0]{kagome2trans2.png}
%(f) \includegraphics[clip,width=5.3 cm,angle=0]{kagome2trans3.png}
\caption{(Color online) Variation of transmission probability with energy for (a) $\alpha = 1.0$ and (b) $\alpha = 4.0$ respectively.}  
\label{dostransport}
\end{figure}
%%%%%%%%%%%%%%%%%%%%%%%%%%%%%%%%%%%%%%%%%%%%%%%%%%%%%%%%%%%
The spectral response of this self-similar fractal model can be generally manifested if we study the transmission characteristics with respect to the energy of the electron. The behavior is examined by taking the peano fractal of an arbitrary generation placed between a pair of semi-infinite ordered leads, named as source and drain, respectively. The variation is plotted in Fig~\ref{dostransport} (upper panel). As we see, the off-diagonal anisotropy has a crucial impact on the single-particle eigenstates. Unequal values of $t_x$ and $t_y$ (when $\alpha \neq 1$) create a non-transmitting region of energy nearly $-5 \leq E \leq 5 .$ This regime is situated in between two conducting zones. This immediately suggests the possibility of SPME, indicating a distinct separation between two different quantum phases. Thus, the essential remark is that this minimal anisotropy factor can make substantial changes in the eigenspectrum. 
This has also been validated from the analysis of the level-spacing distribution, as mentioned in 
the subsequent discussion.

\begin{figure}
	\centering
	(a)\includegraphics[width=0.3\textwidth]{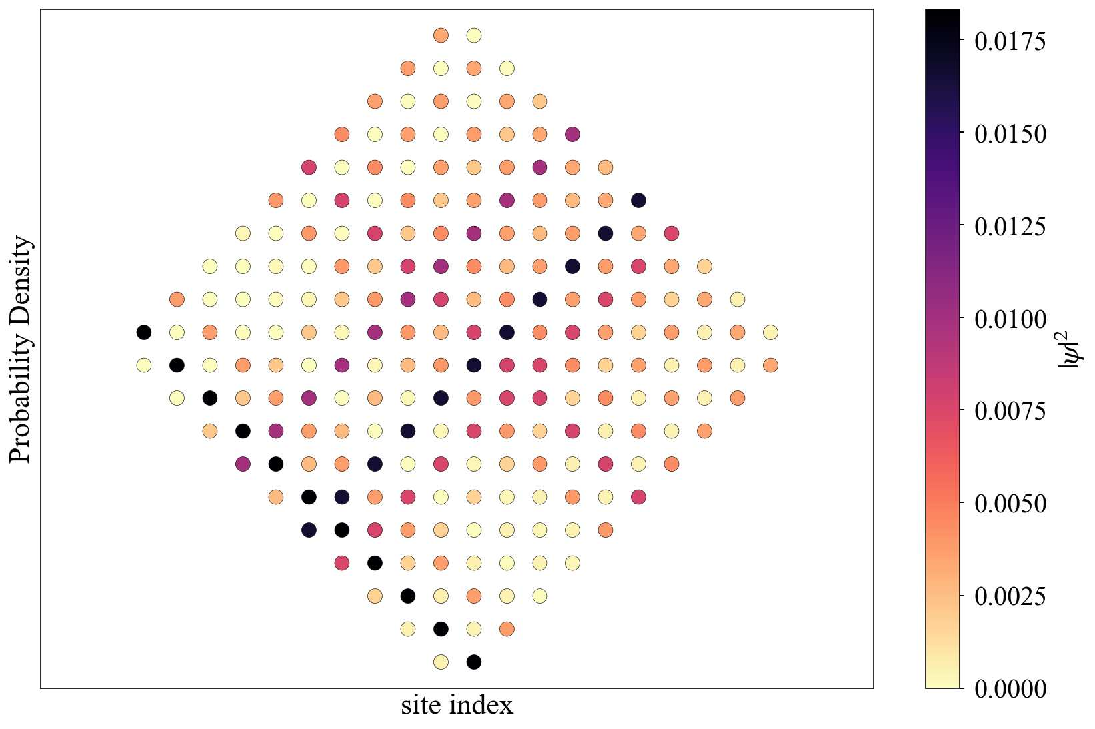}\\
	(b)\includegraphics[width=0.3\textwidth]{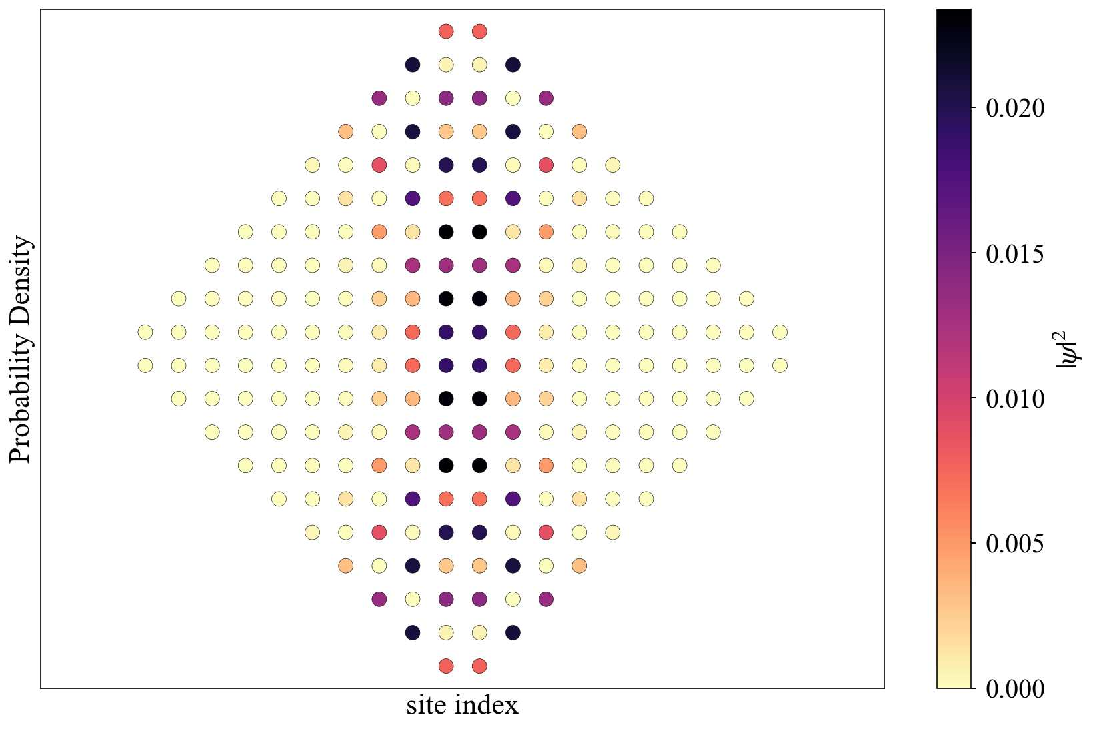}
	\caption{(Color online) Site index probability density heatmap (a) without and (b) with anisotropy $\alpha = 4.0$ to demonstrate the impact of anisotropy on the nature of single-particle eigenstates.}
    \label{siteindex}
\end{figure}
\subsection{Probability density distribution}
From the probability density distribution, as cited in Fig.~\ref {siteindex}, we can notice the contrast between the isotropic and anisotropic hopping configurations for the eigenvalue closest to $E = 0$. This profile indicates that for the isotropic peano fractal, the probability is broadly distributed over the entire lattice, while after introducing off-diagonal anisotropy, the dynamics of the particle is concentrated in a relatively small spatial island in the lattice geometry. These results are consistent with the transport phenomena we observed in this peano fractal lattice.
This demonstrates that hopping anisotropy induces pronounced localization of the eigenstate. The extended state in the isotropic case supports appreciable coherent transport near $E=0$, whereas the localized state in the anisotropic case reduces the spatial overlap between the source and drain. Consequently, the transmission $T(E)$ is strongly suppressed around $E=0$, consistent with the transport spectrum.

\subsection{Level spacing statistics}
\begin{figure}
	\centering
	(a)\includegraphics[width=0.35\textwidth]{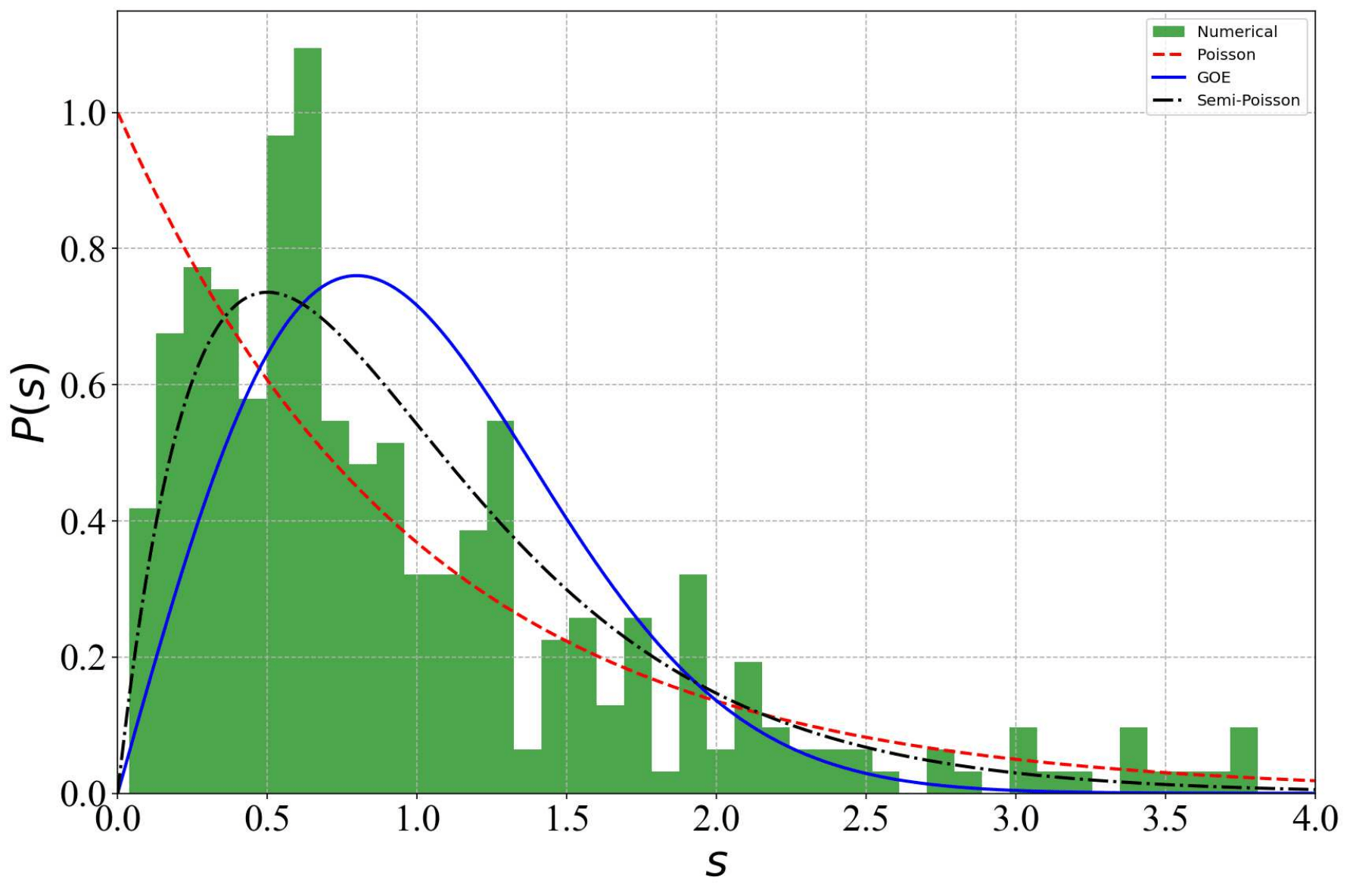}
	(b)\includegraphics[width=0.35\textwidth]{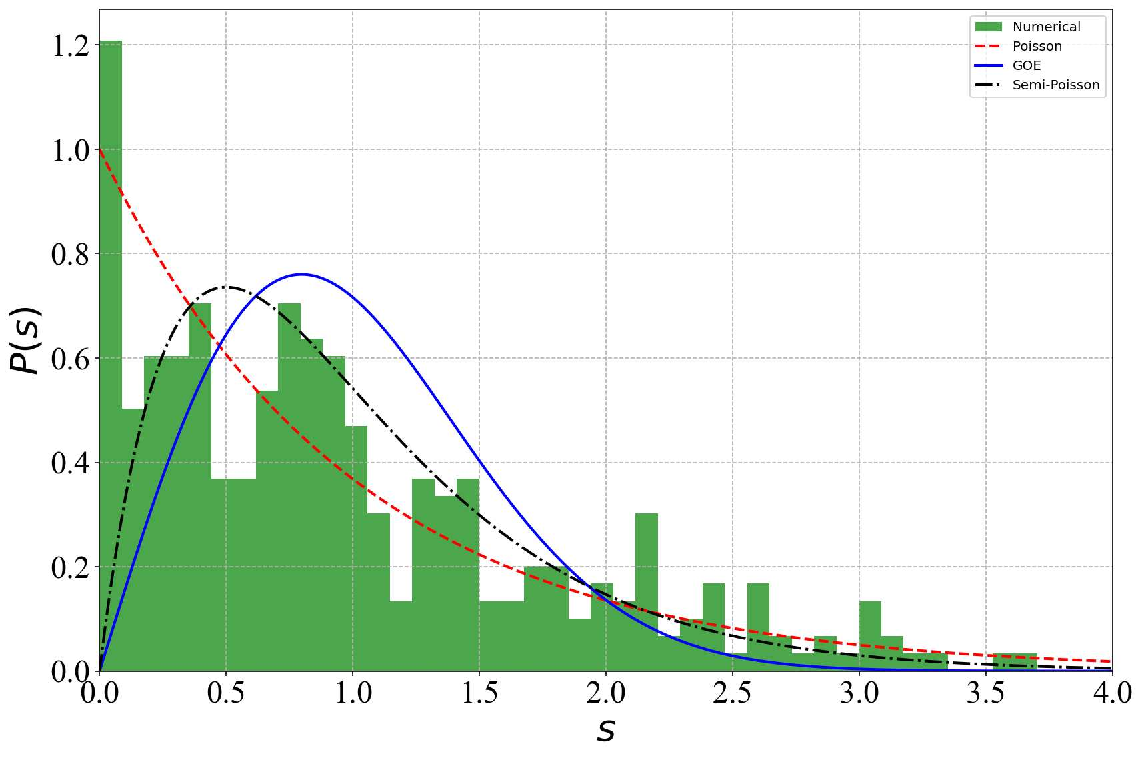}
%	(c)\includegraphics[width=0.35\textwidth]{evsalpha.png}
	\caption{Level spacing distribution in (a) absence and (b) presence of anisotropy $\alpha=4.0$.}
	\label{levelspacing}
\end{figure}

%The subtle control over the transmission profile can be verified from the level spacing distribution.
To justify the above findings, we investigate the nearest-neighbor level spacing distribution of the single-particle eigenspectrum. Level spacing statistics quantify the correlations between adjacent eigenvalues and provide an independent
tool to characterize the imprisonment of electron. Extended states generally exhibit
level repulsion because of strong hybridization between nearby
eigenstates, whereas localized states tend to produce weakly correlated
energy levels and an enhanced probability of small level spacing.
The upper panel of Fig.~\ref{levelspacing} shows the level spacing distribution for the
isotropic fractal.
In this case, the numerical distribution is suppressed near $s=0$ and
develops a broad maximum at a finite normalized spacing. The suppression
of nearly degenerate levels indicates appreciable level repulsion. The
distribution consequently shows a stronger resemblance to the Gaussian orthogonal ensemble (GOE)
prediction than to the monotonically decreasing Poisson distribution.
Although
the distribution does not match exactly with the ideal
GOE form. Such deviations are expected because the peano lattice is a
finite deterministic hierarchical structure rather than a conventional
disordered random-matrix system. Finite-size effects, residual
degeneracies, geometric symmetries, spectral gaps and the coexistence
of states with different spatial characters can all produce departures
from universal Wigner-Dyson statistics. 

Introducing hopping anisotropy produces a remarkable enhancement of the
probability density near $s=0$. At the same time, the finite-spacing
maximum associated with strong level repulsion becomes less pronounced.
The numerical distribution therefore shifts away from the GOE form and
toward a predominantly Poisson-like profile. This crossover indicates
that anisotropy weakens the correlations between nearby energy levels
and reduces the effective hybridization of the corresponding
eigenstates.
It is to be noted that the anisotropic distribution does not follow the ideal
Poisson type throughout the complete spacing interval. A portion of the
numerical histogram remains intermediate between the Poisson and
semi-Poisson curves. This intermediate behavior suggests that the
anisotropic system cannot be described as a uniformly localized phase.
Instead, the spectrum contains states with different degrees of
localization, including weakly correlated localized states, critical
states, and states retaining a finite degree of spatial extension. This indicates a multifractal nature
of the spectrum.

To substantiate the crossover inferred from the level-spacing
distribution, we further consider the average adjacent-gap ratio.
The ratio of two consecutive gaps is then written as,
\begin{equation}
r_n=
\frac{
\min\left(\Delta_n,\Delta_{n+1}\right)
}{
\max\left(\Delta_n,\Delta_{n+1}\right)
}
\end{equation}
and its spectral average is
\begin{equation}
\langle r\rangle
=
\frac{1}{N_r}
\sum_{n=1}^{N_r}r_n
\end{equation}
where $\Delta_n=E_{n+1}-E_n.$ is the successive energy gap.
The gap-ratio statistic does not require spectral unfolding and
therefore provides a useful complementary measure of short-range
spectral correlations. For an uncorrelated spectrum described by
Poisson statistics, the theoretical value is $\langle r\rangle_{\mathrm{P}}
=2\ln 2-1\simeq 0.3863$.
whereas the Gaussian orthogonal ensemble gives $\langle r\rangle_{\mathrm{GOE}}\simeq 0.5359.$
For the isotropic Peano lattice, $\alpha=\frac{t_y}{t_x}=1,$ the average adjacent-gap ratio is found to be approximately $\langle r\rangle_{\alpha=1}\simeq 0.50.$
This value lies close to the GOE benchmark and is consistent with the
pronounced suppression of small level spacings observed in the upper
panel of Fig.~\ref{levelspacing}. The corresponding level repulsion indicates strong
hybridization between neighboring eigenstates and supports the
extended-state character of the isotropic lattice.
In contrast, for the anisotropic case considered here, $\alpha=\frac{t_y}{t_x}=4,$
the representative average gap ratio decreases to $\langle r\rangle_{\alpha=4}\simeq 0.42.$
This value is close to the Poisson benchmark and agrees qualitatively
with the strong enhancement of small spacings in the lower panel of
Fig.~\ref{levelspacing}. The reduction in $\langle r\rangle$ indicates that the energy
levels become weakly correlated when hopping anisotropy is introduced.
Such Poisson-like spectral statistics are characteristic of localized
single-particle eigenstates.
The change $ \langle r\rangle_{\alpha=1} \simeq 0.50 \quad\longrightarrow\quad \langle r\rangle_{\alpha \neq 1} \simeq 0.42 $
therefore signals a clear anisotropy-induced crossover from a
level-repelling spectrum to an approximately uncorrelated spectrum.
Together with the transmission and participation-ratio results, the
gap-ratio analysis supports the existence of extended states in the
isotropic lattice and predominantly localized states in the
anisotropic lattice.

\section{Re-entrant localization-delocalization transition}
\label{transition}
We have shown that a minimal of-diagonal modulation can control the localization of electron in a comprehensive way. The spectral competition between the parameters thus decides the possibility of imprisonment of the excitation.
We now introduce a quasiperiodic modulation following a two-dimensional higher-order AAH model in the off-diagonal term of the Hamiltonian, viz.,$t_x=t_{0x}\left[1+\lambda\cos(2\pi\beta n)\right],$ and $t_y=t_{0y}\left[1+\lambda\cos(2\pi\beta n)\right].$
Here, we choose $\frac{t_{0y}}{t_{0x}} = \alpha$ as before, while the second part gives the modulation factor. The parameter $\lambda$ represents the strength of the modulation, and $\beta$ denotes the periodicity of the modulation. For any rational value of $\beta$, it gives a staggered variation of the overlap integral. However, for any irrational value, $\beta=\frac{\sqrt{5}-1}{2},$ it introduces a quasiperiodic off-diagonal disorder. We shall show that the interplay between the kinetic parameters eventually leads to a re-entrant localization-delocalization transition in the parameter space, accompanied by the possible existence of a single-particle mobility edge (SPME).

\begin{figure}
	\centering
	(a) \includegraphics[width=0.35\textwidth]{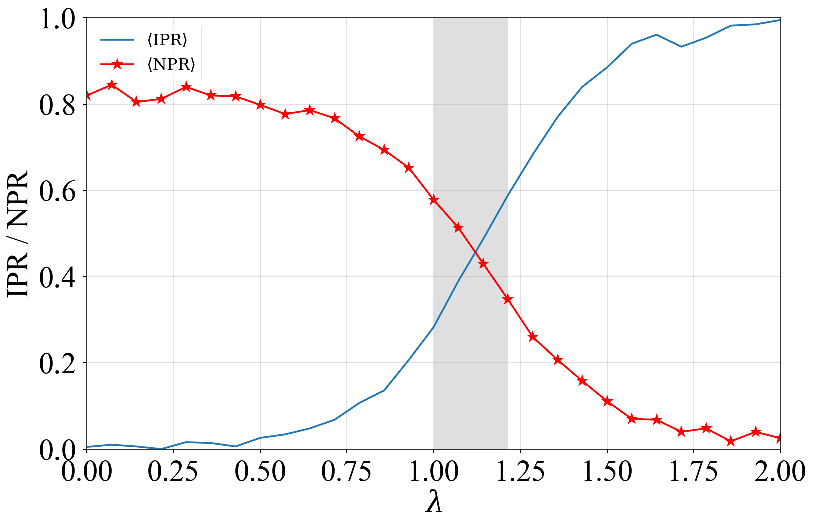}\\
	(b) \includegraphics[width=0.35\textwidth]{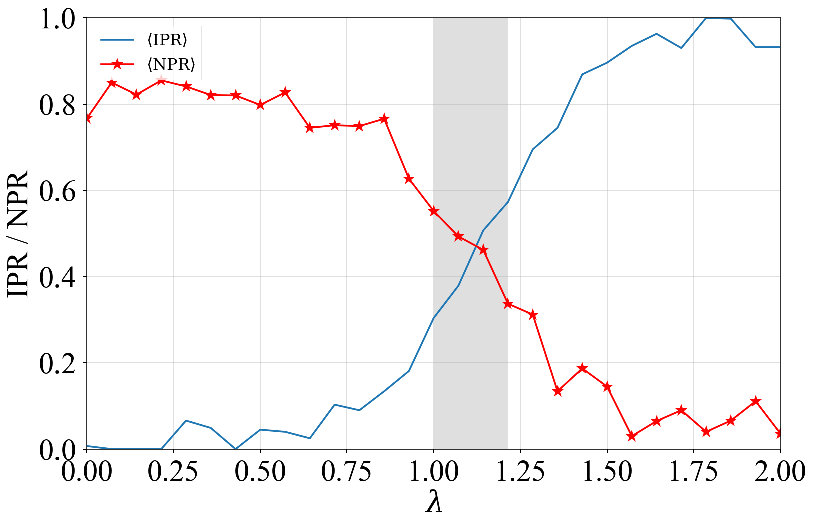}
	\caption{(Color online) Variation of average IPR/NPR with $\lambda$ for (a) $\alpha = 1.0$, (b) $\alpha = 4.0$ respectively.}
	\label{avgipr}

\end{figure}
\subsection{Average IPR vs $\lambda$}

It is well-known that pure AA variation produces a localization transition without SPME. Unlike this canonical case of AA model, we describe that the localization transition happens through a critical regime hosting the SPME for both $\alpha = 1$ and $\alpha \neq 1$. To investigate it, we examine $\langle \mathrm{IPR} \rangle$ and $\langle \mathrm{NPR} \rangle$ as a function of $\lambda$ where the averages are computed by taking all the eigenstates for a certain $\lambda$. As it is seen that the $\langle \mathrm{IPR} \rangle$ and the $\langle \mathrm{NPR} \rangle$ cross each other for $\lambda = 1$, contrary to the original AA case. It is also to be noted that the plots reveal a co-existing sector where both the $\langle \mathrm{IPR} \rangle$ and $\langle \mathrm{NPR} \rangle$ assume finite values. This region is distinctly marked in the Fig.~\ref{avgipr} around $\lambda = 1.0$. This consequently points out the presence of conducting and insulating phase for a range of $\lambda$, $0.75 < \lambda < 1.25$. This is the critical regime of the spectrum associated with the SPME.

\subsection{Localization transition}
\begin{figure}

	(a)\includegraphics[width=0.35\textwidth]{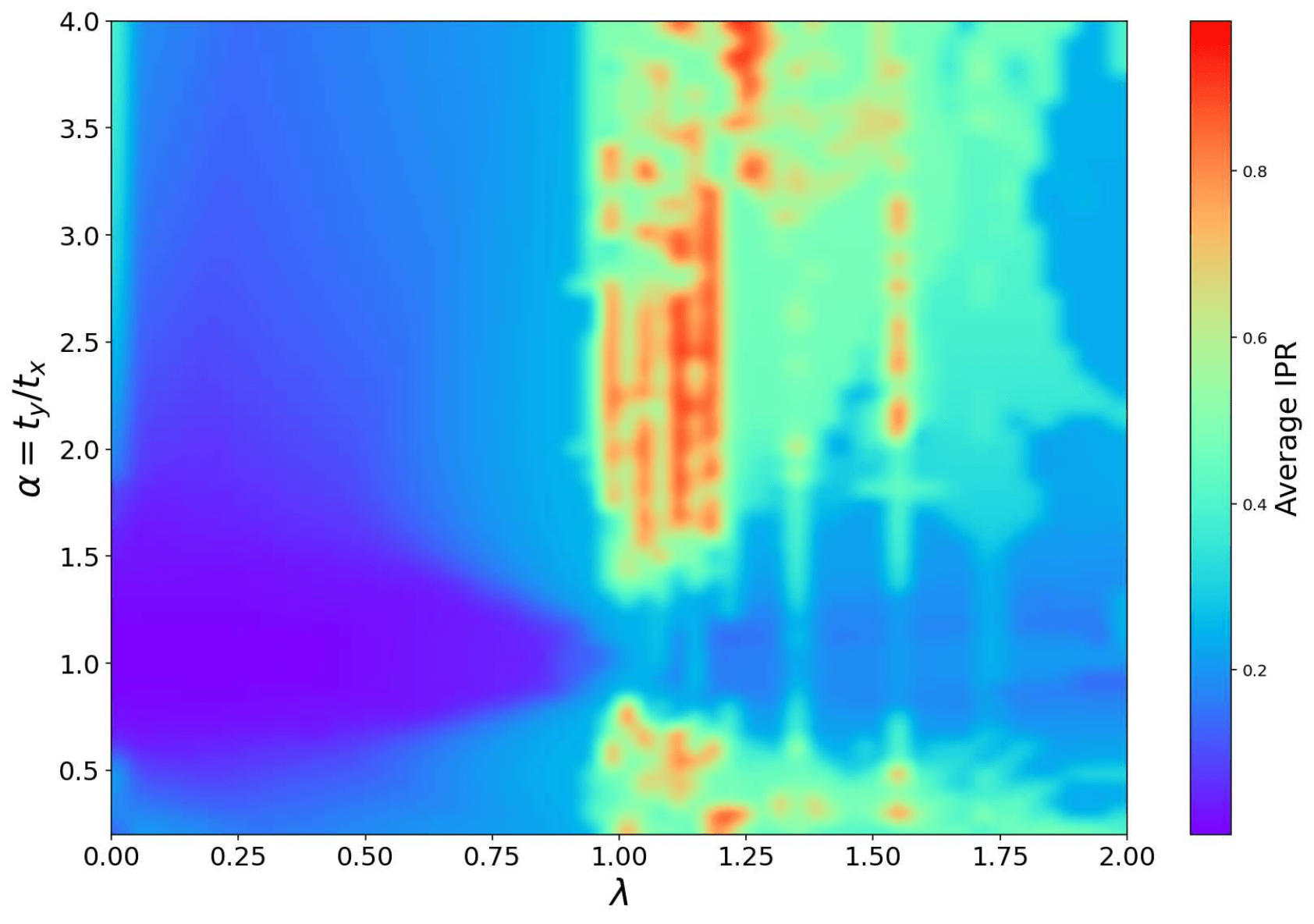}\\
	(b)\includegraphics[width=0.35\textwidth]{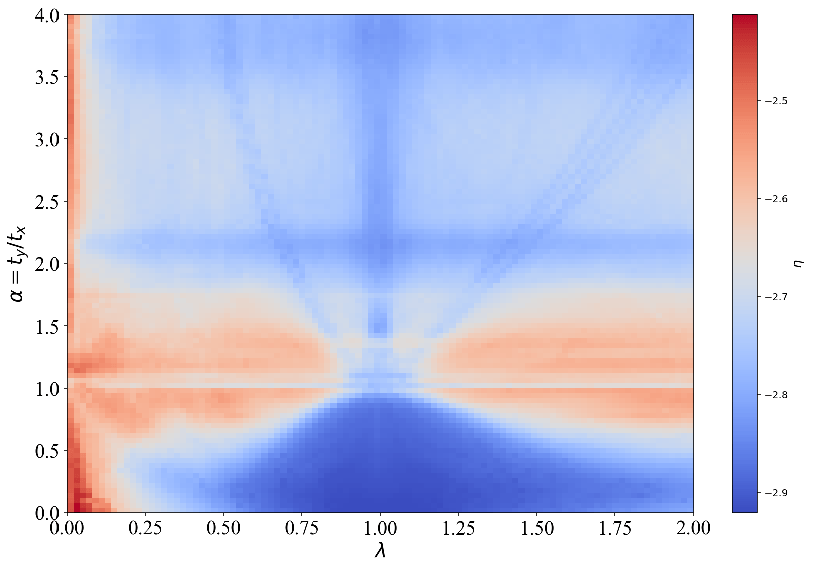}\\
	(c)\includegraphics[width=0.35\textwidth]{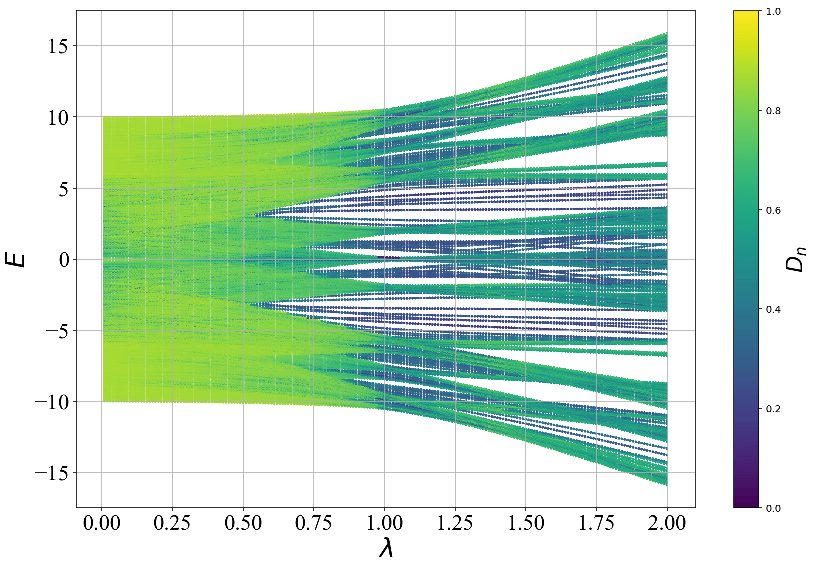}
\caption{(a) IPR intensity heatmap showing the re-entrant localization transition, (b) $\eta$-map and
(c) eigenspectrum with $\lambda$ in respect of fractal dimension.}
\label{hmap}
\end{figure}
The re-emerging localization transition can be easily traced from the IPR intensity plot in the $\alpha$--$\lambda$ parameter space (Fig.~\ref{hmap}). As observed, that the phase plot exhibits a deep blue patch corresponding to extremely low IPR values, indicating the delocalization of the eigenstates. In this region, the competition between the kinetic parameters leads to the diffusion of excitation. For any of the eigenstates, the significant overlap between the wavefunction of the nearest neighboring sites makes the fractal accessible to the incoming electron. As off-diagonal disorder makes its impression on the dynamics of the excitation, the quantum interference effect changes its nature to destructive one.  The system then enters an insulating phase for $\lambda \sim 1$, where the red region signifies the localization of the wave packet. The high value of IPR for the eigenstates essentially demands non-transmittivity of the electron. Thus, this transformation from conducting to insulating phase is extremely sensitive on the parameters selected. As we move beyond this regime, the resonant phase reappears, as indicated by the emergence of the blue patch. It is interesting to point out that, unlike the case of a random arrangement of scatterers, the re-entrant behavior is clearly observed with slowly varying quasiperiodic hoppings.
 With the introduction of quasiperiodic hopping modulation, the spectral competition between the anisotropy index $\alpha$ and the AAH modulation strength $\lambda$ finally leads to a tunable re-entrant localization-delocalization transition.[s das ref]
 
 Furthermore, this essential result can be justified from the phase diagram of $\eta$. The quantity $\eta$ is generally defined as
$$
\eta=\log_{10}\!\left[\langle \mathrm{IPR}\rangle \times \langle \mathrm{NPR}\rangle\right].
$$
Similar to the phase diagram plotted in respect of average IPR, the $\eta$ map Fig~\ref{hmap}(b) also confirms the claiming of re-emerging localization transition. The red colored region is seen to be populated by delocalized eigenfunctions for low values of $\lambda$. The system then encounters an insulating phase marked by blue regime for $\lambda\approx1$. The extended phase again reappears as we go beyond $\lambda=1$. Thus the interplay of anisotropy index $\alpha$ and the strength $\lambda$ of the quasiperiodic modulation essentially supports an exotic eigenspectrum for this scale-invariant structure.

Before ending this demonstration, we look into the the Fig~\ref{hmap}(c)
where the distribution of energy eigenvalues is plotted with $\lambda$ on the basis of fractal dimension $D_n$. The fractal dimension $D_n$ is the slope of the $\ln IPR$ vs $\ln N$ plot. This quantity serves the confirmation of the nature of single particle eigenstates. From this distribution, it is clear that the spectrum validates the localization transition. A subset of eigenstates goes through a delocalization-localization-delocalization crossover. Beyond $\lambda=1$, a portion of the localized states become again extended for this anisotropic peano fractal. The intermediate part also confirms the existence of SPME.

\subsection{Quantum dynamical response}
The imprisonment of the incoming electron can be alternatively visualized if we study the quantum dynamical response~\cite{dyn} of the underlying system. This is a standard strategy to observe the evolution of wavefunction with time. The basic essence in this regard
is that if the electron is localized in space, then the probability of finding the electron at that specific trapping area after a long time will be certainly finite. On the contrary, the probability diminishes with time for a diffusive eigenstate. From this
standpoint, the time-dependent Schr\"{o}dinger's equation helps us to calculate the dynamical property of the system which primarily unfolds the possibility of caging with respect to different parametric configurations in a better way. The controlled kinematics and hence the re-entrant localization-delocalization transition, as discussed earlier, can therefore be easily demonstrated by evaluating the temporal evolution of wavepacket. The essential key evaluating quantity that tells about this fact is the so-called mean-square displacement (MSD).
	\begin{figure}
		\centering
		\includegraphics[width=0.35\textwidth]{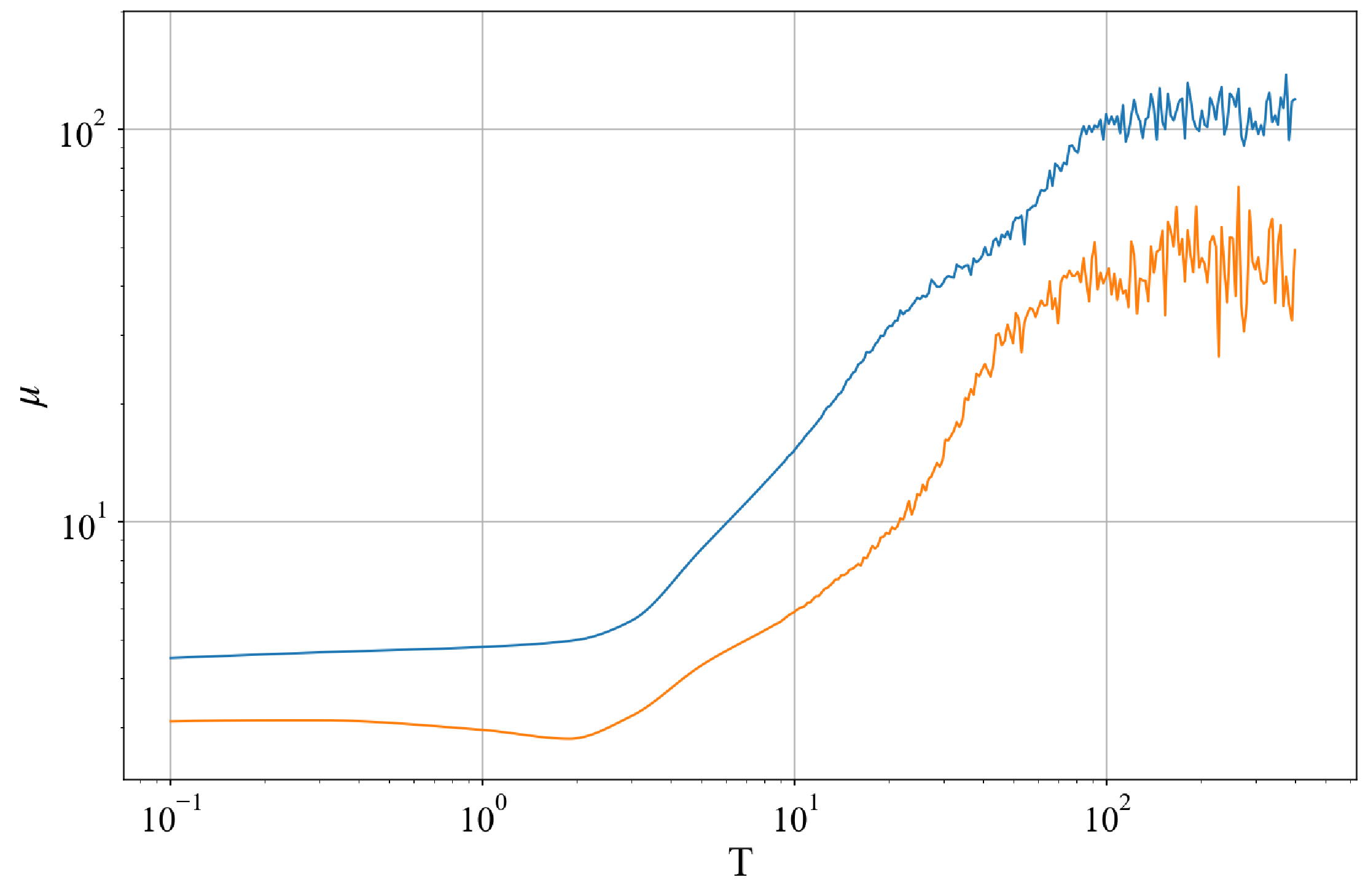}
		
		\caption{(Color online) Temporal variation of dynamical response in absence and presence of
		anisotropy for a $5^{th}$ generation system containing $1860$ number of sites.  }
		\label{msd}
	\end{figure}
	
	In this section, we explore the quantum dynamical properties of the peano fractal under isotropic and anisotropic conditions with different values of AAH amplitude, $\lambda$, configurations. We have taken $4^{th}$ generation peano fractal with number of sites $1860$. The initial position of the wave-packet is $m = 930$.
	
	At time $t = 0$, the particle's wave packet is given by,
	\begin{equation}
		\left|\Psi(0)\right\rangle
		=
		\sum_{n} C_n(0)\left|\psi_n\right\rangle
	\end{equation}
	The initial probability amplitude for finding a particle at the $n^{th}$ site is denoted by $C_n$. When the particle is initially released at the mth site, this amplitude is given by
	\begin{equation}
	C_n(0)
	=
	\langle \psi_n | m \rangle
		\end{equation}
	After some time, the temporal evolution of the state can be represented as,
	\begin{equation}
		\left|\Psi(t)\right\rangle
		=
		\sum_{n} C_n(0)\, e^{-iE_n t}
		\left|\psi_n\right\rangle
	\end{equation}
	The mean square displacement maybe formulated as,
	\begin{equation}
		\sigma^2(t)
		=
		\sum_{n} (n-m)^2 |\psi_n(t)|^2
	\end{equation}
	
	\begin{equation}
		\sigma_n^2(t) \propto t^{\mu},
	\end{equation}
	Where $\mu$ is the diffusion exponent. It is evident that the diffusion exponent $\mu$ characterizes the transport regime: $\mu=1$ corresponds to normal diffusion, $\mu < 1$ to subdiffusion or localized, and $\mu>1$ to $\mu = 2$ superdiffusion (ballistic or enhanced transport).
	
	First we identified the extended and localized region from the Fig.~\ref{hmap} which is $\lambda = 0.50$ \& $\lambda = 1.25$ respectively. With these $\lambda$ values we have computed the MSD with time $t$. Fig.~\ref{msd} provides the evolution of the MSD for both the cases. We fixed the hopping anisotropy to $\alpha = 4.0$. When $\lambda = 0.50$ the scaling exponent is $\mu=1.75$ which lies in the super-diffusive regime. This behavior is consistent with the heat map provided earlier, indicating that the wave packet efficiently explores a large portion of the peano network before saturation due to finite size of the fractal. In contrast, when $\lambda = 1.25$, the spreading of the wave packet is strongly suppressed. The MSD saturates much earlier, reflecting the confinement of the particle around its initial position. This observation agrees with the scaling exponent $\mu=0.41$, which belongs to the localized regime. The reduced spatial spreading originates from the AAH amplitudes $\lambda$ and hopping parameters. With the change of hopping anisotropy $\alpha$ and AAH amplitudes $\lambda$ we can thus establish that the system exhibit the re-entrant localization transition from this dynamical perspective.

\section{Quantum butterfly and edge modes}
\label{butterfly}
\subsection{Multifractal butterfly spectrum}
It is in general advisable to present the allowed energy eigenvalues with respect to the variation of the frequency of the imposed AAH modulation. The spectrum is cited in the Fig~\ref{butter}. It reveals the formation of a quantum butterfly pattern which is an interesting variant of Hofstadter butterfly indicating a multifractal landscape. A close observation unfolds interesting crossover and opening of disallowed regimes which ultimately finalizes this eigenspectrum. Non-trivial change of band curvature with $\beta$ is also noticed in this quasi-continuous spectrum. The gaps provide a setting for the appearance of boundary-localized states. In our system, edge modes are situated at the wings of the butterfly. Their presence indicates that the quasiperiodic hopping does not only fragment the bulk spectrum but also generates nontrivial boundary states. Thus, the quantum butterfly provides a direct spectral signature of the relation between AAH modulation and edge-state formation in the system.

\subsection{Robust edge modes}
Edge states are electronic states whose wave functions are predominantly localized near the boundaries of a finite-sized system. Unlike bulk states, these states exhibit a large probability density at the boundary while remaining weakly distributed in the interior of the lattice. The formation and robustness of such boundary states depend on the lattice geometry, hopping amplitudes, and external perturbations. In system like fractals, AAH modulated hopping amplitudes can generate localized edge modes through quantum interference.

To characterize the nature of the eigenstates, we examine the probability density $|\psi_i|^2$ of selected eigenstates corresponding to representative energies in the butterfly spectrum. The degree of boundary localization is quantified by the edge probability

\begin{equation}
	P_{\mathrm{edge}}=\sum_{i\in \mathrm{edge}} |\psi_i|^2,
\end{equation}

where the summation extends over all boundary sites of the lattice. The quantity $P_{\mathrm{edge}}$ represents the edge weight of an eigenstate. States with large values of $P_{\mathrm{edge}}$ are identified as edge-localized states, whereas those with small values correspond to bulk states.

\begin{figure}

	\includegraphics[width=0.31\textwidth]{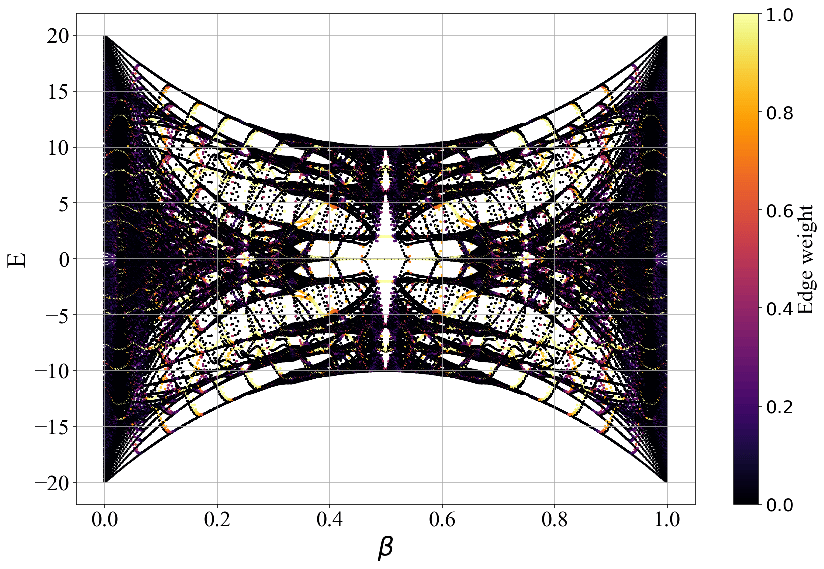}
	\includegraphics[width=0.31\textwidth]{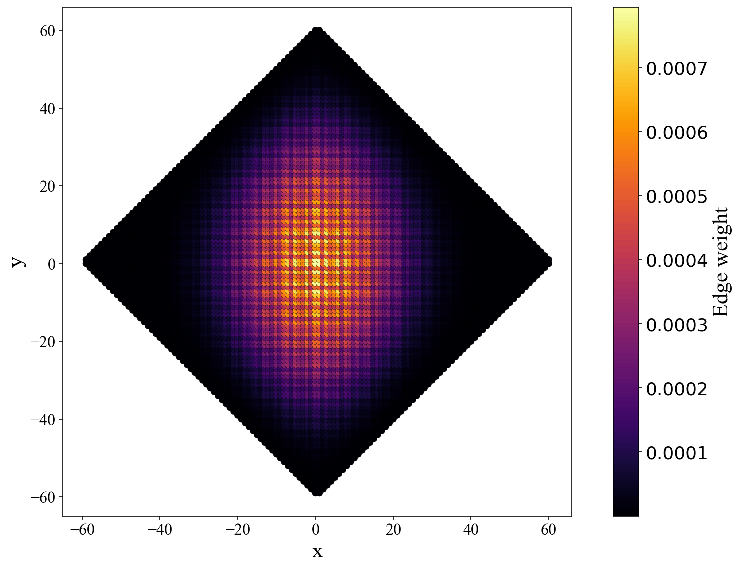}
	\includegraphics[width=0.31\textwidth]{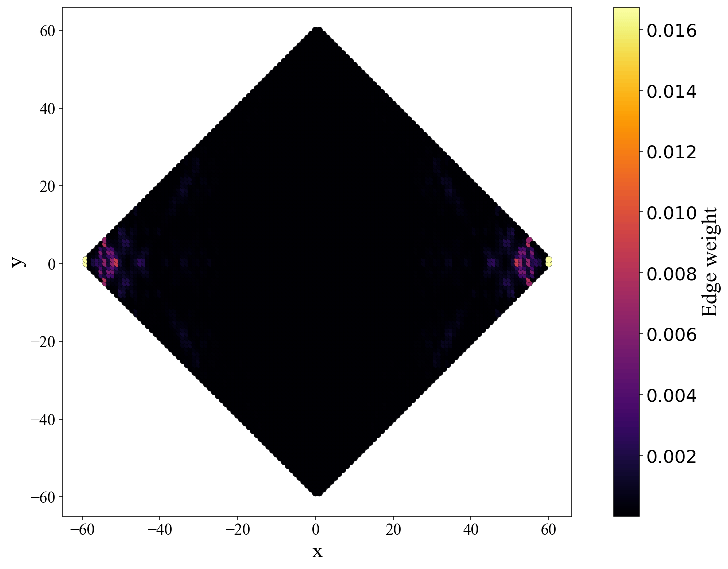}
	
	\caption{(Color online) (a) Quantum butterfly spectrum with the edge probability mapped onto each eigenstate. Bright yellow regions correspond to states with large edge probability, while dark regions represent bulk states.(b) Probability density of a representative bulk state $E = 0.35$. (c) Probability density of a representative edge-localized state $E = 3.75$. }
	\label{butter}
\end{figure}

To identify boundary-localized states, we first compute the edge probability $P_{\mathrm{edge}}$ for every eigenstate and map it onto the quantum butterfly spectrum, as shown in Fig.~\ref{butter} (a). Bright yellow regions correspond to eigenstates with large edge probability, whereas the dark branches represent bulk states. Representative eigenstates are then selected from these two distinct regions, and their real-space probability densities are analyzed. The resulting wavefunction profiles clearly distinguish edge-localized and bulk states, as illustrated sequentially in Fig.~\ref{butter}(b) and (c).

Our numerical results reveal the existence of boundary-localized eigenstates whose probability density is strongly concentrated near the boundary of the peano fractal while remaining negligible in the interior. In contrast, bulk states are distributed throughout the lattice. These observations demonstrate that the quasiperiodic AAH hopping modulation supports the emergence of localized boundary modes.

To examine their robustness, random disorder is introduced into the quasiperiodic hopping amplitudes according to

\begin{equation}
	\lambda_i \rightarrow \lambda_i+\delta_i,
\end{equation}

where $\delta_i$ is uniformly distributed in the interval $[-W_d/2,W_d/2]$. The edge modes remain localized upto a critical disorder strength of approximately $W_d^{c}\approx 3.0 $, beyond which the boundary localization gradually weakens because of increasing hybridization with the bulk states.

Furthermore, we numerically verified that the Hamiltonian preserves chiral symmetry. The persistence of the edge-localized states under moderate disorder is consistent with the preservation of this symmetry, which maintains the spectral symmetry of the system and contributes to the stability of the boundary modes against those perturbations.
\begin{figure}
	(a)\includegraphics[width=0.3\textwidth]{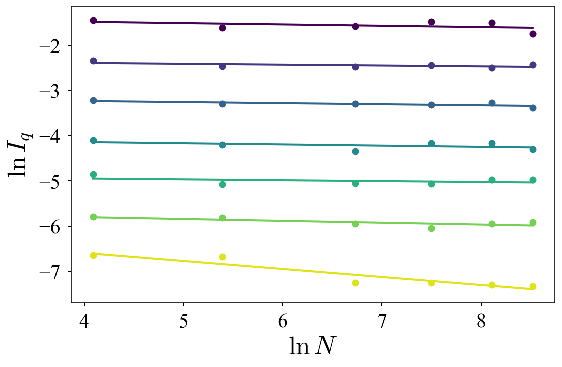}\\
	(b)\includegraphics[width=0.3\textwidth]{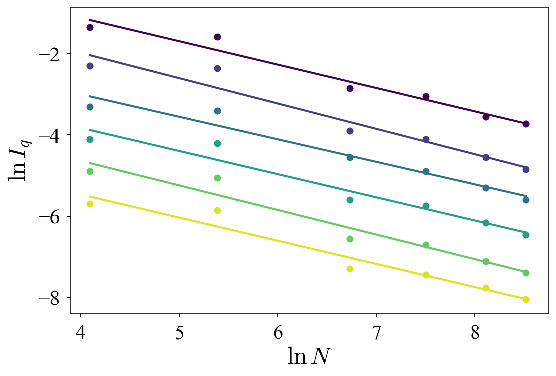}\\
	(c)\includegraphics[width=0.28\textwidth]{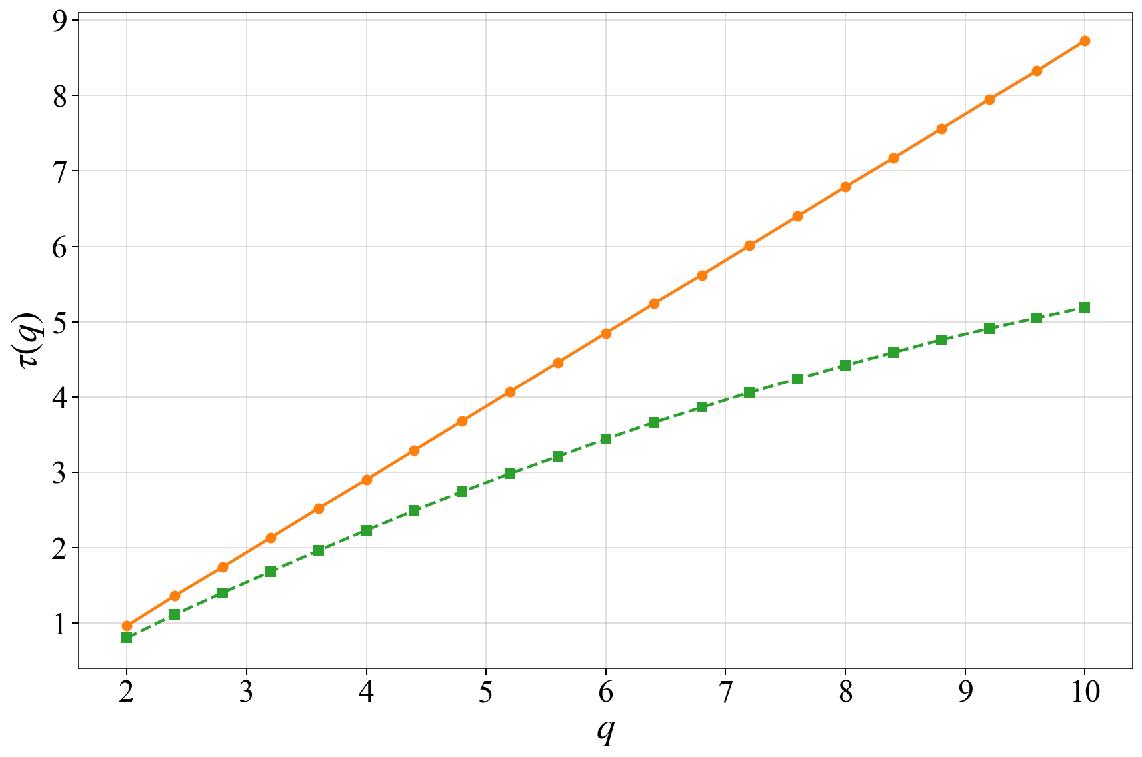}
	\caption{(Color online) GIPR profiles with system size for
	(a) $E = 1.30$ and (b) $E = 2.50$ and (c) $\tau(q)$ plot showing the multifractality.}
	\label{mfa}
\end{figure}
\subsection{Multifractality and mobility edge}

We define the SPME as a critical energy threshold which separates insulating phase from the non-localized modes in a quantum system. In one-dimensional system with uncorrelated disorder, there cannot be any SPME as all the states localize exponentially even in presence of weak or moderate disorder. However, for quasiperiodic systems the system can host a localization transition leading to a metal-insulator transition with appropriate parametric correlation. The IPR plot with energy for a certain generation of the system as presented in the Fig.~\ref{mfa} provides the preliminary evidence of concurrence of two different quantum phases. The IPR value shows a remarkable jump and it eventually becomes a gapless eigenspectrum for $\beta=\frac{\sqrt{5}-1}{2}$, a Diophantine number. We have set the anisotropy index $\alpha=4$. To do this we have carefully chosen two suitable eigenvalues, viz., $E=1.3$ and $E=2.5$ respectively with low and high IPR values. To differentiate physically between the two eigenstates, we therefore take the help of generalized inverse participation ratio (GIPR) in terms of the moment of wavefunction amplitude $|\psi(r)|^2$ as,
\begin{equation}
I_q=\int |\psi(r)|^{2q},d^d r
\end{equation}
GIPR follows a standard power-law style with the system size $L$,
\begin{equation}
I_q\propto L^{-\tau(q)}.
\end{equation}
The exponent $\tau(q)$ is thus defined as,
\begin{equation}
\tau(q)=-\lim_{L\rightarrow\infty}\left(\frac{\ln I_q}{\ln L}\right).
\end{equation}
The generalized fractal dimension $D_q$ which is related to multifractal exponent $\tau(q)$ as,
\begin{equation}
\tau(q)=D_q(q-1)
\end{equation}
It characterizes the multifractal signature of the spectrum. $\tau(q)$ maintains linearity with $D_q=d$ for conducting phase and becomes a nearly flat with $D_q=0$ indicating the insulating phase. It shows a typical nonlinear behavior with $q$ at critical points.

In accordance with this standard prescription addressed above, the two distinct quantum phases coexist in the eigenspectrum. For $E=1.3$, the $\tau(q)-q$ plot is a signature of extended mode and non-linear $\tau(q)-q$ variation is denoting the localized mode for $E=2.5$. It is to be mentioned that the sustainability of the states has been checked for larger generation of the fractal. The nonlinear plot of $\tau(q)-q$ for $E=2.5$ is an inherent feature of multifractality of the allowed eigenspectrum.

	\section{conclusion}
	\label{conclude}
	In this article, we have initially investigated a typical response of a self-similar peano fractal with suitable kinetic modulation. Minimal off-diagonal anisotropy is seen to degrade the transparency of the system to the incoming electron resulting in a central non-conducting window in the energy landscape with respect to the modulation factor. The results are justified by the evaluation of level-spacing distribution and transmission co-efficient. The non-trivial change with nominal kinetic anisotropy induces us to undertake the same fractal with quasiperiodic variation in the overlap integral. By analyzing the average IPR and NPR it has been concluded that the system undergoes a re-entrant localization-delocalization transition hosting a SPME. Our claim has been confirmed through the critical regime identified with respect to the strength of modulation. The entire analysis is corroborated with the standard $\alpha-\lambda$ phase plot and $\eta$-map. Moreover, the formation of quantum butterfly pattern along with few in-gap corner states are recorded within the tight-binding framework. These corner states are robust and found to be protected by the chiral symmetry associated with the underlying system. Quasiperiodic modulation imposed on a fractal substrate can thus present exotic energy spectrum including re-entrant phase transition under certain correlation of parameters.
%%%%%%%%%%%%%%%%%%%%%%%%%%%%%%%%%%%%%%%%%%%%%%%%%%%
%\begin{acknowledgments}
%AN is thankful for the stimulating
%discussions regarding the results with Prof. Arunava Chakrabrti.
%\end{acknowledgments} 
%%%%%%%%%%%%%%%%%%%%%%%%%%%%%%%%%%%%%%%%%%%%%%%%%%%

%%%%%%%%%%%%%%%%%%%%%%%%%%%%%%%%%%%%%%%%%%%%%%%%%%%
\end{document}